\documentclass{sig-alternate-05-2015}

\usepackage{graphicx}
\usepackage{array}
\usepackage{xcolor}
\usepackage{cleveref} 


\begin{document}

\setcopyright{rightsretained}


%


\title{Interactive Lab Notebooks for Robotics Researchers}

%
%
%
%
%

%
\numberofauthors{1}

\author{
\alignauthor Rolando Garcia
    \affaddr{UC Berkeley}
    \email{rogarcia@berkeley.edu}
}

\maketitle
\begin{abstract}

Interactive notebooks, such as Jupyter, have revolutionized the field of data science by providing an integrated environment for data, code, and documentation. However, their adoption by robotics researchers and model developers has been limited. This study investigates the logging and record-keeping practices of robotics researchers, drawing parallels to the pre-interactive notebook era of data science. Through interviews with robotics researchers, we identified the reliance on diverse and often incompatible tools for managing experimental data, leading to challenges in reproducibility and data traceability. Our findings reveal that robotics researchers can benefit from a specialized version of interactive notebooks that supports comprehensive data entry, continuous context capture, and agile data staging. We propose extending interactive notebooks to better serve the needs of robotics researchers by integrating features akin to traditional lab notebooks. This adaptation aims to enhance the organization, analysis, and reproducibility of experimental data in robotics, fostering a more streamlined and efficient research workflow.

\end{abstract}

\section{Introduction}

Data scientists are increasingly reliant on, and skilled with, interactive notebooks like Jupyter
for modeling and analysis. Prior to interactive notebooks, data scientists worked with a loose
conglomeration of programming languages, shells, statistical analysis software, visualization
tools, and so on~\cite{kandel2012enterprise, shankar2024we}. This past heterogeneity and frequent hand-off made it hard for scientists to
keep their code, data, metadata, plots, and thoughts organized beyond a publication deadline.
As time passed and memory faded, data scientists would occasionally find themselves getting
stuck in a reproducibility eﬀort, or feeling confused about the underlying reasons for outliers
and anomalies. Interactive notebooks unified such disparate solutions into a singular
environment — data scientists could use interactive notebooks to view, transform, annotate,
and display data all together, on the same page. By grouping data, code, plots, and plain-
English explanations in a single document, interactive notebooks made it possible (but far from
guaranteed~\cite{head2019managing,kery2018interactions}) for data scientists to trace the lineage of a plot back to the code and data that
produced it, just by scrolling up on a web page. Today, interactive notebooks are a lively and
rapidly growing area of research.

And yet, model developers — especially those who work with neural networks — as well as
robotics researchers, have resisted interactive notebooks for all but the most basic
demonstration use cases~\cite{amershi2019software}. To many, such disinterest is perfectly reasonable: model
developers do not use training data in a scientific capacity; they use it instrumentally and only
for evaluation purposes. Swap a model developers training data and they may grumble about
tuning; swap a scientists data and --- no, better not to interfere with a scientist's data. 
In this
sense, a model developer is no more a data scientist than a mathematics professor (who uses
linear algebra for image compression) is a photographer.

But there is a sense in which model developers (and robotics researchers) are data scientists:
when they generate, capture, organize, transform, and plot data about their training
experiments for publication. This is data about how long the model took to finish training, and
how well it performed on tests. It's data about how many times the model developer iterated
on designs, with what configuration and with what results. Model developers act in a scientific
capacity when they design experiments, run those experiments, record observations, and
analyze or synthesize the resulting data. And, when model developers or robotics researchers
act in a scientific capacity, we find that they make the same mistakes as data scientists made
before the introduction of interactive notebooks.

We did not set out to answer the question, ``why don't model developers use interactive
notebooks?'' Because we, like most, took it for granted that it didn't make sense for model
developers to use them. Rather, we set out to answer a diﬀerent question, ``how do model
developers track experiment data, how do they manage records, and so on?'' What we learned
was that their workflow was remarkably similar to the dominant workflow of data scientists
before interactive notebook were introduced~\cite{amershi2019software}. We found that robotics researchers use a
diverse, even disparate set of tools for their scientific process; they link a large set of code,
artifacts, and results in working memory, which is suﬃcient for the production of a journal article, but not a durable solution. We found that robotics researchers are either too hesitant to
modify legacy code written by graduate students from many years ago, because the students
are no longer around and they're afraid they could break it. Alternatively, they may repeatedly
build and re-build one-oﬀ tools and workflow management systems for each experiment.

\subsection{Contributions}

In this paper, we discuss a philosophical framework through which we analyze the interview data and consider implications for system designs. The framework emerged after multiple cycles of coding and refinement. Our theory is that a robotics researcher is a data scientists just as much as is a computational biologist. We propose an alternative use of \textit{data science} with the goal of clarifying the proper scope of data science tools such as interactive notebooks. With this
framework, we are driven to the conclusion that interactive notebooks get a lot right, but need to
be extended so that they behave more like traditional (i.e. pen-and-paper) lab notebooks.

Our initial hypothesis was that robotics researchers logged fewer physical parameters than was
necessary. This didn't really turn out to be the case. Rather, they log as much information as
they typically need for the current experiment, but they do a poor job of linking the information
digitally, so with the passage of time, the data can get lost.

We also underestimated the extent to which robotics labs are controlled environments.
Through trial-and-error, robotics researchers have learned to control lighting conditions, noise
levels, foot traﬃc, and so on. A highly controlled environment is good news for the systems
designer, it means that the tool can leverage information from its surroundings to provide its
service.

After discussing observations from our interview study, and covering findings from our thematic
analysis, we explore implications for system design. To summarize our findings succinctly, it's a
shame that model developers and robotics researchers disregard interactive notebooks,
because these could help them manage their scientific workflow more easily. Rather than
extending interactive notebooks to fit their needs, robotics researchers spend their eﬀorts on
workarounds that increasingly start to resemble existing tools. The good news is that we can
quickly extend interactive notebooks to meet the existing needs.

\section{Theoretical Framework}

I share the constructivist view that words and their meanings impact much more than our
social dynamics: they have the capacity to shape our tools and their uses in a manner that
aﬀects our material reality. What we mean by data science directly influences what we expect
when we hear about a new data science tool. As we negotiate and then converge on a
meaning of data science, we develop expectations for what a good data science tool should
do. When our use of data science is too narrow, the tools built for data science are too limited.
We see this with interactive notebooks, which \textbf{lack even rudimentary support for data entry},
despite the fact that recording data and measurements during an experiment is a crucial piece
of the scientific process. In this section, I will make the case that we should adjust our use of
data science to mean something slightly diﬀerent; something more, that respects established
precedent but is also more consistent with what we mean by science. With this new meaning,
we will see that interactive notebooks have a more important role to play in the scientific
process, beyond mere data analysis. Interactive notebooks have the capacity to do for
experiment design and data entry what they have done for data analysis: \textbf{consolidation and
rendezvous of disparate parts}. 
In a future section we will discuss the promises of such an
extended interactive notebook for scientists and data scientists alike.

Before continuing further, it's worth pausing briefly to ask ourselves, what for do scientists use
traditional lab notebooks? what for do scientists (or data scientists) use interactive notebooks?
And, can traditional lab notebooks and interactive notebooks be used interchangeably for the
same thing? Scientists work meticulously with traditional lab notebooks, because these serve
as authoritative records of discovery in patent disputes, as proof of work for funding agencies,
and as step-by-step instructions for reproducibility. Scientists extensively use traditional lab
notebooks throughout these four activities:
\begin{enumerate}
    \item The design of the experiment.
    \item Observation arising from the experiment.
    \item Logging or journaling of parameters and measurements.
    \item The analysis or synthesis of data from experiments.
\end{enumerate}
Data scientists, on the other hand, use interactive notebooks for (and only for) the fourth
activity: data analysis. They are able to get by despite such limited support because they often
work with data created by someone else, scientist or not. Because interactive notebooks are
data science tools, and data science in the standard sense has such a narrow scope, it's no
surprise that interactive notebooks lack even the most rudimentary support for data entry and
recording of observations from experiments.

We propose thinking of data science more broadly, and shifting the nexus of interactive notebooks
to the traditional lab notebook and away from the interactive textbook (i.e. literate programming~\cite{perez2015project}). In this new framing,
data science is everything a scientist does that typically requires the use of a computer and a
lab notebook. Specifically, activities 1-4 enumerated above. Data science thus conceived
elucidates important limitations in existing interactive notebooks (i.e. lacking support for
activities 1-3), and illustrates directions for future work, such as the unification of computers
and scientific lab notebooks. With the help of this framework, we can articulate why model
developers and robotics researchers, need, but don't presently want, next-generation
interactive notebooks. The need follows from the fact that they are at present enduring similar
problems of provenance and context collapse as data analysts endured in the past (pre-2014).
The disregard follows from the fact that current-generation interactive notebooks are based on
the literate programming paradigm (i.e. interactive textbooks) instead of the lab notebook
paradigm, and are therefore a poor fit for experiment design and data entry.

\section{Methods}

\begin{table*}[t]
    \centering
    \begin{tabular}{| l | l | l | p{4.5cm} | p{4.5cm} |}
        \hline
        \textbf{ID} & \textbf{Affiliation} & \textbf{Title} & \textbf{Area} & \textbf{Robot} \\ \hline
        P0 & UC Berkeley & EECS faculty & robotics, machine learning, reinforcement learning, systems & Sensor network, swarm \\ \hline
        P1 & UC Berkeley & EECS PhD Student & robotics, reinforcement learning & Surgery robot, laparoscopy \\ \hline
        P2 & UC Berkeley & EECS PhD Student & robotics, reinforcement learning & Warehouse arm \\ \hline
    \end{tabular}
    \caption{List of participants interviewed about their logging or record-keeping practices in robotics research.}
    \label{tab:recruitment}
\end{table*}

Following recent systems work for logging, or record-keeping, in machine learning~\cite{Garcia2020} we had
observed that model developers followed very agile practices, which at times could seem ad-
hoc or lacking in discipline. We found that the explanation for this behavior was the low-cost of
missing training records: most times model developers could finish their task with light or low
logging; in cases when they needed more details, they could add more logging statements to
their program and re-run. We also knew that these logging practices diﬀered substantially from
the careful and meticulous record-keeping of experimental scientists. We then wondered what
a middle-ground for logging methodology may look like. Following this line of inquiry, we found
that robotics researchers appear to be balanced between model developers and experimental
scientists: they train digital models on advanced computers for AI tasks, but the robot's
behavior and the researcher's observations are grounded in real-world phenomena, much like
the more classic experimental scientist. Perhaps more interesting was the fact that if robotics
researchers intentionally or accidentally repeated the logging behavior of model developers,
that they would find themselves in trouble: there is no way to rewind physical time to more
carefully observe real-world phenomena. Do robotics researchers know this? Have they
adapted? Are they preventive or reactive?

We began our study with very little knowledge about robotics, so it was diﬃcult to theorize
early or ask leading questions. There were some positives to this open-mindedness: our lack of
background forced us to stick to simple and concrete questions instead of more theoretical
and general questions. With mixed success, we were able to see some problems or events
through the eyes of the participant: we could imagine being there. So much novelty also had its
drawbacks: a lack of a suﬃciently relevant theory early in the study made it diﬃcult for us to
notice surprises or contradictions. A lack of theory also meant an inability to diﬀerentiate
between signal and noise during an interview, and from the transcripts it appears that we could
sometimes chase irrelevant leads or stay near the surface with basic clarification questions. We
learned the hard way that qualitative data is nearly impossible to interpret in a theory vacuum.
So, although we began our study without a theory, our theory began to emerge after multiple
passes over the interview data.

We responded to the challenge of information overload (in the early stages of the study, when
we lacked focus) by inviting other members of the research group to be active listeners and
occasional participants in the interview. These additional members completed the required
training and were approved by IRB prior to the interviews. This increased our ability to catch
and process total information, but it also meant the injection of occasional tangents and
distractions: a consequence of diﬀerent people having diﬀerent interests and backgrounds.
In the end, we were left with a plurality of blurry perspectives instead of one clear view. In order
to render a coherent conceptual scene, we turned to Grounded Theory~\cite{muller2014curiosity}.

We are indebted to
Grounded Theory for our theoretical framework (i.e. interactive notebooks as lab
notebooks and robotics researchers as data scientists), thematic analysis, and implications for
system design. Grounded Theory method gave us the necessary flexibility to theorize after
data collection by doing numerous in-depth passes over the interview transcripts, codes, and
memos; refining, and iterating. Although the work was labor-intensive, it made an otherwise intractable
analysis possible.

\begin{figure*}
    \centering
    \includegraphics[width=0.9\linewidth]{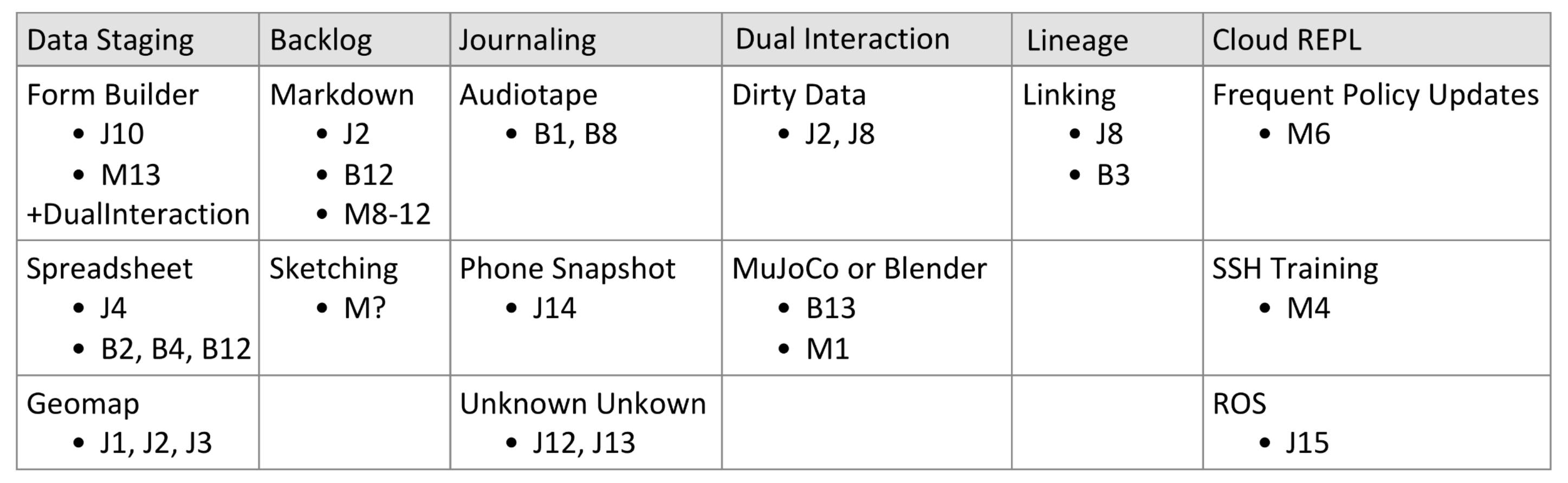}
    \caption{Data matrix of themes in interview transcripts. Each element in a cell
corresponds to a theme, the bulleted codes correspond to transcript block tags.
Themes are grouped by next-gen interactive notebook components.}
    \label{fig:codes}
\end{figure*}

\subsection{Recruitment}

We conducted three interviews of robotics researchers at UC Berkeley (\Cref{tab:recruitment}). Participants were identified through UC Berkeley robotics labs websites, and pre-screened by navigating to their professional websites. Following approval from UC Berkeley's Institutional Review Board, we cold-emailed four participants to invite them to participate in
our interview study, and three responded aﬃrmatively and signed written consent forms; the
other did not respond. We emailed candidates in batches to allow ourselves time to refine the
codes, theory, and protocol interview between batches. Participants were each compensated \$60.00 in the form of VISA gift cards, except for P0 who was not compensated.

\subsection{Semi-structured Interview}

Our semi-structured interview followed an arc of kick-oﬀ, building rapport, grand tour, and
reflection. The structure is typical in need finding studies because it pads the in-depth interview
with questions that the interviewee is able to answer with little eﬀort and thus get settled in and
get comfortable. The goal of the grand tour is to prompt the interviewee, by our active listening
and asking a series of incrementally deepening questions, to give us a walk-through or tour of
an applicable memory. Better to see one memory through their eyes than to hear
generalizations which are meaningful to them but empty to us. We wish to learn from their past
as they have learned from their own experience.

In order to expedite IRB approval,
we limited our interviews to no more than two hours, did not schedule follow-up interviews,
and did not record audio or video of our call -- as required by the strict guidelines for exempt
status. All interviews were conducted over Zoom, and transcripts were automatically
generated by the service.

\subsection{Transcript Coding and Post-Processing}

After each interview, we post-processed the transcripts, added comments, and added codes.
A few days after each interview, the study personnel met to exchange ideas and insights, and
ask questions. We jointly wrote memos to summarize and supplement the transcripts.
Transcript post-processing consisted of redacting personally identifiable information,
attributing transcript segments to personnel versus participant, and correcting some errors
made by the transcription service. Although laborious, we used the post-processing pass as an
opportunity to carefully read and understand the interview, and add comments. Coding was
performed in subsequent passes.

Once we finished post-processing, commenting, and coding all interviews, we still felt lost in
the data and the patterns were elusive. One strategy in Grounded Theory is to build a
new theory from existing coding scheme(s). The theory and codes evolve together, informing
one another, until reaching equilibrium. In the end, the resulting coding scheme contains
implicit information about the emerging theory. The analyst can then return to the interview
transcripts and perform an additional coding pass. In our case, we found that new codes
continued to emerge. When there were new codes we wanted to add, we would think
about how to adapt the theory to accommodate them, and would consider the gaps in the
theory that led to our omission of these codes. We repeated this refinement process until the coding scheme and theory converged. In a sense, this set of methods helped us fit the coding
scheme and theory to our set of interviews. We know that to establish generalizability and
validate our theory, we will have to conduct future interviews and measure how well our coding
scheme and theory fit previously unseen data. A rule of thumb is to continue repeating this
process until subsequent interviews stop surprising us, and the coding scheme and
accompanying theory stabilize.

\subsection{Thematic Analysis}

After coding the transcripts, we performed thematic analysis on our data, the findings of which
we present in the next section. Thematic analysis is a qualitative analysis method for identifying
patterns or recurring themes in the data~\cite{hornof2017designing}. These themes help us organize our theory into
manageable groups, and support the theory with concrete examples from the data. For
thematic analysis, we scanned the coded transcripts for blocks of text that had left a lasting
impression. By now, following multiple coding passes, the authors had developed succinct
overviews of each interview in working-memory. Thematic analysis is thus a means for us to
externalize internal representations we developed by working with the data. By now, we could
recall details of each transcript by first remembering a series of highlights. Thus, doing a pass
over the transcripts to find blocks of code that were characteristic of themes (or highlights) was
straightforward. Once these blocks of text were extracted, a new labeling pass was done to
characterize the blocks of text. 

\section{Summary of Findings}

We conducted our thematic analysis after settling on a theoretical framework (described in
section 2). For convenience, we restate our theory here: robotics researchers are data
scientists who design experiments, run those experiments, record data and measurements
from experimental observations, and analyze or synthesize their experimental data for reporting
or publication. Robotics researchers need powerful interactive notebooks that have more in
common with scientific lab notebooks than with digital textbooks. We find that the state-of-the-
art in robotics research resembles the scattered and heterogeneous world of computational
data analysis before the introduction of Jupyter notebooks, where code, data, lineage, and
other digital information is linked for the duration of a publication sprint, but as time passes
and memory fades, context degrades and scientific results become irreproducible;
experimental observations become uninterpretable.

\subsection{Data Staging, Backlogs, and Journaling: Jobs for
Scientific Lab Notebooks}

In this section we describe data entry techniques that have emerged in robotics research, and
which can be managed with technologies as simple as pen-and-paper. After discussing some
least bounds on functionality for data entry with notebooks, we discuss data entry practices
that robotics researchers rely on, and opportunities for tool development.

\subsubsection{Data Staging} Staging is a technique for agile data entry that postpones data integrity checks until later. In
computing, data that is staged is entered tentatively, and may be subsequently committed into
a permanent record upon passing review. An analogy for staging is penciling in some
measurement on a clipboard; the measurement is committed into the record when written in
pen. Staging enables the robotics researchers to both observe experimental phenomena
online, and validate data entry oﬄine. Tooling should ensure that suﬃcient context survives into
the future so that researchers can focus on experimental phenomena in real-time, knowing that
they can reasonably evaluate the quality of staged data after-the-fact.

P0 illustrates the process of moving staged data to increasingly more durable and structured
stores:
\begin{quote}
    And so I needed to convert the notes that I wrote in a piece of paper, which was like,
``sensor 6 was placed on [a near] desk; sensor 7 was placed on [a far] desk, on the
other side of the lab \dots'' And I would draw like roughly where in the building [the motes
were placed], like, ``the left corner of that kitchen.'' And then I would have to get a floor
plan, figure out the coordinates in the floor plan, map the pixel space, and then I would
translate the sensors to those [coordinates], so I could use a statistical model to figure
out, for every pixel \dots what the Wi-Fi might have been, had I put a sensor there to build
a new model of the building.
\end{quote}

Later P0 shares an example of what committing staged data would look like :
\begin{quote}
    As soon as I deployed the sensor, I would go straight to the floor plan and try to read it.
I would transcribe all my notes onto the map, the floor plan, and I would actually like,
get the pixel location, and I'd make a spreadsheet which is like, ``the x and y coordinate
of that sensor in the picture of the floor plan.'' So that from that on I could just use the
picture the floor plan and not my notes.
\end{quote}

Today, robotics researchers stage experimental data mainly through spreadsheets. We identify
the following tooling opportunities for data staging:
\begin{itemize}
    \item Continuous context capture to enable post-hoc validation and cleaning of staged data.
    \item Form building utilities to prompt, guide, or remind the researcher about what data to enter.
\end{itemize}

P1 discussed their own solution for continuously capturing context for data staging:
\begin{quote}
    So we are logging basically all the trajectory data, essentially to like a giant pickle file.
But we’re also like recording videos, and also we have a spreadsheet where we’re
recording timing information as well.
\end{quote}

\subsubsection{Backlog (Sprint Planning)}
The backlog is like a todo-list for robotics researchers, and it regards the Python code that
powers the robot’s computers, such as the policy or the perception model. When the
researcher observes undesirable behavior, they may at times speculate about the reasons why
there is a failure. Instead of carrying out the fix right away, researchers may take down some
notes and return to the problem later, e.g., to allow the current experiment to finish. In another
scenario, perhaps because the robotics researcher is ready to end their work day, they may
write down some notes to help themselves resume work in the morning. In the words of P2:
\begin{quote}
    Yeah, I think people, people will take notes, fairly often, I mean, I keep a notebook that I
just like write whatever I feel like, not the most organized thing but uh just kind of like
thoughts \dots I think some of it is like to go back and look at later right like you're just
writing down something of this is more important than something else.
\end{quote}
There are some similarities between the backlog and the data stage: both serve as temporary
placeholders for work that is too important to ignore but not so important that it merits doing
immediately; both add technical debt and rely on context; both communicate some degree of
promise and risk about the entry; and either may be asked to drop an element from its basket if
that element fails to meet some quality standard.

There are also some important diﬀerences between the backlog and the data stage. When
committed, staged data enters a permanent record, such as a database, a log, a chart in a publication, and so on. When committed, or more aptly completed, an entry in the backlog
ends in patching some outstanding issue in the robot’s code.

Based on our interviews, we believe that robotics researchers manage their backlog with pen-
and-paper. They may take down a note in a paper notebook, draw a sketch, or write a
symbolic equation. They manage context in natural or intuitive ways, such as managing spatial
context incidentally by entering a note on a post-it note and attaching the note to a robot in a
frozen state.

P2 shares his story about how they manage their backlog in a paper notebook. Note how
entries that relate to diﬀerent projects or diﬀerent responsibilities are co-mingled in the same
notebook:
\begin{quote}
    So if I'm working on like derivations or something, and see how this *points at
notebook* can be like random math that I've written down. There will be like this *points
at page* A lot of this *points at separate page*, so I mean, I have weekly meetings with
my undergrads, so you can see like \dots Yeah, so a lot of these are like from [my meetings
with my undergrads] \dots I don't know if you can really see it but they'll just be like, kind
of notes that I've taken throughout the meeting of, you know \dots here’s what we're
discussing, what we're working on next week, all this.
\end{quote}

From skimming his own notes, P2 was sometimes confused about what some of them were
about. This may not be a problem if the notes are residuals of backlog entries that made their
way into a patch or a Wiki, but we do wonder about how many backlog entries get lost in the
weeds. Symbolic equations are particularly prone to losing context.
\begin{quote}
    a lot of times I'll just write down like the \dots like in this case I like box some experiment,
basically that I did and like talked about, maybe what could be, what could be wrong
with that one \dots Honestly, like some of these notes are fairly old so I don't actually
remember what that was about but there's yeah there's plenty of stuﬀ like that in here
as well.
\end{quote}

\subsubsection{Journaling}
Journaling is best when it happens continuously and exhaustively. Scientists are taught early to
record observations, the more the merrier, even if they appear to be noise or common-place.
Observations should be logged especially when we don't know what to do with them. The
assumption is that we may presently lack crucial information to notice anything interesting in
the observation, or that the observation may seem irrelevant at this moment simply because
we are working on a diﬀerent experiment and suﬀering from tunnel vision. Were we to learn
something new or work on something else, this same observation could help us solve a puzzle.

P0 motivates journaling quite beautifully in one story. He tells us of a time when he deployed a
sensor network in a football field so there wouldn’t be any interference from the walls or other
electronic devices. P0 was trying to measure a baseline condition. He was surprised to find
that the signal strength was weaker in the football field than in a campus building. The reason,
unbeknownst to him at the time, was that ground muzzles radios, and you should never put
radios on the ground. What metadata would P0 have to log for an independent 3rd party to be
able to diagnose the problem from afar? That the motes make direct contact with the ground
would do the trick, if the 3rd party has enough background in physics. But even if we grant
that, if P0 does not have the requisite knowledge, it will not even occur to them to log this
information, the surface where a sensor is placed doesn’t even register as metadata to track.
So what’s another way that P0 could send enough information to the 3rd party to get help
diagnosing the problem? P0 could send a photo of the deployment:
\begin{quote}
    I don't know that I have a good way to index [photographs] but if I were to publish that
data, in the long run, that might have been very useful because someone else might go
``why is a signal so crappy for this football field deployment?'' \dots ``Oh, they put [the motes] in the grass, who does that??'' they'd actually have the information. Whereas, they
wouldn't when I did it the way I did it also like when I put a mote on a desk like was it a
metal desk? sometimes. That might have been a bad idea so like they're probably
important details that could have helped [to record].
They might have actually aﬀected the conclusions \dots
And that might have been a saving. Like nowadays, when I go and I need to remember
something I just take a picture like I go to a hotel and I remember like the login code in
my room number I just take a picture of the key card and I'm done.
\end{quote}
We then asked future interview participants how they felt about logging with video. Although
there were no objections, P1 expressed some skepticism about how complete video logs
could be. After all, some relevant metadata does not have a visual representation.

P1 told us of a time when one of the surgery robot's internal cables was not engaging correctly,
and this manifested itself in the robot not moving. If all you are recording is video, one could
think of dozens of reasons why the robot is not moving. To diagnose the problem P1 moved
the robot's wrist around and listened carefully to the sound of clicking. Although the wrist
would click, the sound was not as per usual. He then strummed cables inside the robot's
forearms and noticed that the tension was also not as per usual. What kind of journaling could
P1 use to help future undergrads diagnose this particular problem? One idea is to follow a ``think
aloud protocol'', in which you say everything that is crossing your mind, what you think the
problem is, what you are feeling and listening for, what surprises you, and so on. I asked P1
whether he was explaining his actions to the undergrad he was with, and he said that indeed
he was:
\begin{quote}
    Yeah, so there was actually one more grad student as well there too, so yeah, I was
trying to, I was trying to explain what was wrong.
I didn't like properly verify whether they understood what I said. But, um, yeah I was like
talking about what I was doing and why I was doing it, but it's it's very hard to
understand unless you do it yourself, so I'm not like so sure that I was very clear. But I
was talking about what I was doing and why I was doing it, it looked really weird \dots like
it just looked like I was like feeling up the robot, but I was like doing something very
specific.
\end{quote}

So, although a 3rd party listening to an audio tape of P1 describing his diagnostics process
may not be able to follow what is happening, it's possible that an undergraduate student, when
encountering a similar problem could try to repeat the same motions, and synchronize the
voice tape with her own actions for guidance. In summary, audio and video are promising
technologies for continuous and exhaustive logging of robotics experimental setups.

\section{Reflections}

Grounded Theory was the right methodology for analyzing our set of interviews, given that we
started the study with very little background and without a theory. Grounded Theory imposed a
heavy labor burden at analysis time, but it helped make an otherwise intractable problem of
distilling a theory and identifying patterns something manageable.

Constructivism was an important philosophical device for helping us explaining how, from a
muddied understanding of a word, data scientist, it could happen that people with valid use
cases for interactive notebooks could manage to overlook them for so long.

We would like to validate the theory (Section 2) and findings (Section 4) by implementing a
solution and testing it. One advantage of engineering research (even qualitative engineering
research) is that it's possible to take the theory and insights, and use them to build a tool for
some identified need. The extent to which the tool is eﬀective is a proxy for the quality of the
methodology, the validity of the findings, and the plausibility of the theory. In the short run
though, we would like to continue running interviews, because we are still very far from the
saturation point, and expect future interviews to be full of surprises.

\section{Acknowledgements}
This report was submitted in partial fulfillment of EDUC 271B, \textit{Introduction to Qualitative Research Methods}, a graduate course at UC Berkeley taught by Professor Erin Murphy-Graham. We would like to thank Bobby Yan and Melissa Salazar, who were undergraduate researchers at the time who were involved in the interviews. 

\bibliographystyle{abbrv}
\bibliography{sigproc}  

\end{document}